\documentclass[11pt]{article} 
\usepackage{amsmath} 
\usepackage{amssymb} 
\usepackage{graphicx} 
\usepackage{geometry}
\usepackage{cite}

\def\bsigma{\mbox{\boldmath $\sigma$}}
\def\bxi{\mbox{\boldmath $\xi$}}

\begin{document}

\title{{\bf Multiplicative versus additive noise in multi-state neural
networks}}

\author{D. Boll\'e$^{a,}$\footnote{desire.bolle@fys.kuleuven.ac.be},
J. Busquets Blanco$^{a,}$\footnote{jordi.busquets@fys.kuleuven.ac.be}, 
T. Verbeiren$^{a,}$\footnote{toni.verbeiren@fys.kuleuven.ac.be}\\
$^a$Instituut voor Theoretische Fysica, Katholieke Universiteit
  Leuven\\
Celestijnenlaan 200D, B-3001 Leuven, BELGIUM}

\date{}
\maketitle

\begin{abstract}
The effects of a variable amount of random dilution of the synaptic 
couplings in $Q$-Ising multi-state neural networks with  Hebbian 
learning are examined. A fraction of the couplings is explicitly 
allowed to be
anti-Hebbian. Random dilution represents the dying or pruning 
of synapses and, hence, a static disruption of the learning process
which can be considered as a form of multiplicative noise in the 
learning rule.
Both parallel and sequential updating of the neurons can be treated.
Symmetric dilution in the statics of the network is studied using
the mean-field theory approach of statistical mechanics. General
dilution, including asymmetric pruning of the couplings, is examined
using the generating functional (path integral) approach of  disordered
systems. It is shown  that random dilution 
acts as additive gaussian noise in the Hebbian learning rule with a mean
zero and a variance depending on the connectivity of the network and on 
the symmetry. Furthermore, a scaling factor appears that essentially 
measures the average amount of anti-Hebbian couplings.
\end{abstract}

\section{Introduction}

In general, artificial neural networks have been widely applied to memorize 
and retrieve information. During the last number of years there has been
considerable interest in neural networks with multi-state neurons using the
framework of statistical mechanics, which deals with large systems of
stochastically interacting microscopic elements (see, e.g., \cite{B04} and
references cited therein). In these models, the firing states of the neurons
or their membrane potentials are the microscopic dynamical variables. 
Basically, compared to models with two-state (=binary) neurons, such models
can function as associative memories for grey-toned or colored patterns 
\cite{TI01,IC01} and/or allow for  a more complicated internal 
structure of the recall process, e.g., a  distinction between the exact 
location and the details of a picture in  pattern recognition and the 
analogous problem of retrieval focusing in the framework of 
cognitive neuroscience \cite{SNK98}, a combination of information
retrieval based on skills and based on specific facts or data 
\cite{KI93,S86}.

Different types of multi-state neurons can be distinguished according to 
the symmetry of the interactions between the different states. Here we
are primarily interested in the so-called $Q$-Ising neuron, the states of
which can be represented by  scalars, and the interaction between two
neurons can then be written as a function of the product of these scalars. 
So, the $Q$-states of the neuron can be ordered like a ladder between 
a minimum and a maximum value, usually taken to be $-1$ and $+1$.
Special cases are $Q=2$, i.e., the well-known Hopfield model 
\cite{Hopfielda} and  $Q=\infty$, i.e., networks with analogue or graded 
response neurons \cite{Hopfieldb,MW89,MWW90}.

In analogy to the Hopfield model, the multi-state neuron models we
discuss here have their immediate counterpart in random magnetic systems 
(=spin-glasses) (cfr., e.g., \cite{SK} and \cite{GS}), but with couplings 
defined in terms of embedded patterns through a learning
rule. Since one of the aims of these networks is to find back the
embedded patterns as attractors of the recall process, they are also
interesting from the point of view of dynamical systems.

This close relation with spin-glass systems means that the methods and 
techniques used to study the latter have been successfully applied to
these network models. In particular, it also means that concepts like 
temperature, fluctuations, disorder, {\em noise}, stochasticity \ldots  
play a crucial role. In the literature it is well-known 
(see, e.g.,~\cite{GOS99}) that noise 
can have rather surprising and counterintuitive effects in the behavior
of dynamical systems. It has been shown many times that noise can have a
constructive rather than a destructive role. Relevant examples~\cite{GOS99}
 of this
fact are the phenomenon of stochastic resonance, noise induced ordering 
transitions, noise induced disordering phase transitions and an increase
of the maximal information content with dilution 
in some  neural network models~\cite{BEVp}. In principle,
these ordering effects seem to be related to the multiplicative character
of the fluctuations, as compared to the disordering role of additive 
fluctuations. But things are not so simple because there is an interplay
between additive and multiplicative noise terms.  

Moreover, there are several types of internal fluctuations, e.g., 
thermal fluctuations introduced through a random term, quite often assumed
to be gaussian distributed with zero mean and uncorrelated at different times.
These internal fluctuations are described by an additive noise term,
i.e., a random term that does not depend on the variable under
consideration. This is not a necessary character of internal
fluctuations. In some systems these can also be described by a
multiplicative noise which is coupled to the state of the system. As a
natural extension of the concept of internal fluctuations, external
noise are those fluctuations that are not of thermal origin.

In the $Q$-Ising neural network models we are considering here, the following 
noise can be characterized. First, the neurons are stochastic such that
the analogue of temperature is introduced. This enables us, as hinted at
already above, to use the techniques of statistical mechanical mean-field 
theory, and ultimately to compute, e.g., the storage capacity of the network
\cite{B04}. The
zero-temperature limit will always reduce our system to a deterministic 
Hopfield or multi-state $Q$-Ising network. 
The meaning of this stochastic behavior is to model that neurons fire
with variable strength, that there are delays in synapses, that there
are random fluctuations in  the release of transmitters \ldots. 
Briefly, we model this 
internal noise  by thermal fluctuations.

Secondly, in single pattern recall with many -- a fraction of the size of
the network -- stored patterns there is the generally nontrivial interference
noise due to the other patterns. This noise has been treated, e.g.,  by 
statistical neurodynamics \cite{B04,AM,Nishimori} or 
functional integration methods \cite{C01,SMR73,Dom78,BEVp}.

Thirdly, in the case of the Hopfield model the Hebb learning rule has
been generalized in \cite{Som86} in order to bring the model closer to
natural systems. In particular, two types of noise terms have been
added. The first one, an additive external contribution which is
independent of the learning algorithm, and assumed to be gaussian
distributed, is relaxing the hypothesis that the entire synaptic
efficacy is coming from the learning process. The second one, a
random multiplicative factor of order ${\cal O}(1)$, represents
a static disruption of the learning process. An  important example of
the latter is random dilution of the network by the pruning or
dying of synapses, relaxing the unrealistic
condition that every neuron is connected to every other one.
The effects of both these static fluctuations on
the recall process in the Hopfield model have been estimated using
equilibrium  mean-field theory statistical mechanics. Technically
speaking, the use of this method allows for symmetric dilution only, because 
the detailed balance principle, i.e., absence of microscopic probability
currents in the stationary state,  is needed to define an energy function.
An additional  remark is that non-linear updating of the synapses
is allowed in that work. It has been shown that all these 
effects can be represented by an additive static gaussian noise in
the learning rule and
that the model is robust against the interference of this static noise. 

In this contribution we extend the work of Sompolinsky \cite{Som86} in 
different directions.  Since we use both replica
mean-field theory equilibrium methods and non-equilibrium  functional
integration techniques, the assumption of symmetric couplings is 
not required such that we can treat all forms of dilution. Moreover, we
allow a fraction of the couplings to be anti-Hebbian \cite{HFP83}. We 
can also have sequential and parallel updating of the neurons, 
and we examine the effect of these noise terms in $Q$-Ising networks.
The main results are that for the forms of dilution we have examined the 
effects can be represented by additive noise in the learning rule and a
scaling factor proportional to the average amount of anti-Hebbian 
couplings. The diluted networks are robust under these effects.

The rest of this paper is organized as follows. In Section 2 we introduce
the $Q$-Ising neural network model and the types of dilution we are 
interested in. Section 3 treats the statics of the model, where detailed 
balance requires, as we will explain, symmetric  dilution. In Section 4 
the dynamics of the network is studied allowing for a general form of 
dilution. Section 5 presents some concluding remarks.

\section{$Q$-Ising networks with variable dilution}
\label{sec2}
Consider a neural network  consisting of $N$ neurons which can take
values $\sigma_i, i=1, \ldots ,N$ from a discrete set
        $ {\cal S} = \lbrace -1 = s_1 < s_2 < \ldots < s_Q
                = +1 \rbrace $.
The $p$ patterns to be stored in this network are supposed to
be a collection of independent and identically distributed random
variables (i.i.d.r.v.), $\{{\xi}_i^\mu \in {\cal S}\}$,
$\mu =1,\ldots,p$,
with zero mean, $\left\langle\xi_i^\mu\right\rangle=0$, and variance 
$A=\left\langle(\xi_i^\mu)^2\right\rangle$. The
latter is a measure for the activity of the patterns. We remark that for
simplicity we have taken these variables $s_k, k=1, \ldots, Q$ 
equidistant and we have also taken the patterns and the neurons out of 
the same set of variables, but this is no essential restriction. 
Given the configuration
        ${\bsigma}_N(t)\equiv\{\sigma_j(t)\},
        j=1,\ldots,N$,
the local field in neuron $i$ equals
\begin{equation}
        \label{eq:h}
        h_i({\bsigma}_{N}(t))=
                \sum_{j=1}^N J_{ij}(t)\sigma_j(t)
\end{equation}
with $J_{ij}$ the synaptic coupling from neuron $j$ to neuron $i$.

All neurons are updated sequentially or in parallel through the spin-flip
dynamics defined by the transition probabilities
\begin{equation}
      \Pr \{\sigma_i(t+1) = s_k \in {\cal S} | \bsigma_{N}(t) \}
        =
        \frac
        {\exp [- \beta \epsilon_i(s_k|\bsigma_{N }(t))]}
        {\sum_{s \in {\cal S}} \exp [- \beta \epsilon_i
                                   (s|\bsigma_{N }(t))]}\,.
\label{eq:trans}
\end{equation}
The configuration  ${\bsigma}_{N}(t=0)$ is chosen as input.
Here the energy potential $\epsilon_i[s|{\bsigma}_{N}(t)]$
is defined by 
\begin{equation}
        \epsilon_i[s|{\bsigma}_{N}(t)]=
                -\frac{1}{2}[h_i({\bsigma}_{N}(t))s-bs^2]
         \,, \label{eq:energy}
\end{equation}
where $b>0$ is the gain parameter of the system.
The zero temperature limit $T=\beta^{-1} \rightarrow 0$ of this dynamics
is given by the updating rule
\begin{equation}
        \label{eq:enpot}
        \sigma_i(t)\rightarrow\sigma_i(t+1)=s_k 
         \qquad\mbox{such that} \qquad
                \min_{s\in{\cal S}} \epsilon_i[s|{\bsigma}_{N}(t)]
            =\epsilon_i[s_k|{\bsigma}_{N }(t)]
\,.
\end{equation}
This updating rule (\ref{eq:enpot}) is equivalent to using a gain 
function $\mbox{g}_b(\cdot)$,
\begin{eqnarray}
        \label{eq:gain}
        \sigma_i(t+1) &  =   &
               \mbox{g}_b(h_i(\bsigma_N(t)))
                  \nonumber      \\
               \mbox{g}_b(x) &\equiv& \sum_{k=1}^Qs_k
                        \left[\theta\left(b(s_{k+1}+s_k)-x\right)-
                              \theta\left(b(s_k+s_{k-1})-x\right)
                        \right]
\end{eqnarray}
with $s_0\equiv -\infty$ and $s_{Q+1}\equiv +\infty$ and $\theta(\cdot)$ the
Heaviside function. For finite $Q$,
this gain function $\mbox{g}_b(\cdot)$ looks like a staircase with $Q$ 
steps. The gain parameter $b$ controls the average slope of 
$\mbox{g}_b(\cdot)$ and, hence, suppresses or enhances the role of the
states around zero.

It is clear that the $J_{ij}$ explicitly depend on the architecture.
We are interested in architectures with variable dilution and we also
want to allow a  fraction of the couplings to be anti-Hebbian. We
realize this by choosing  the couplings according to the Hebb rule 
multiplied with a factor $c_{ij}$
\begin{equation}
     J_{ij}^{c}=\frac{c_{ij}}{c} J_{ij}, \quad 
              J_{ij}= \frac{1}{NA}\sum_{\mu =1}^p \xi_i^\mu \xi_j^\mu
        \quad \forall i,j       \,,
        \label{eq:JD}  
\end{equation}
with the $\{c_{ij}=0,\pm1\}, i,j =1, \ldots,N $ chosen to be i.i.d.r.v and
obeying, in general, a distribution of the form
\begin{equation}
   \textrm{P}[c_{ij}=x]=
         c_1\delta_{x,1}+c_2\delta_{x,-1}+(1-c_1-c_2)\delta_{x,0}
    \label{ND}
\end{equation}
with $c=c_1+c_2=\langle
|c_{ij}|\rangle=\langle c_{ij}^2\rangle$ the connectivity, i.e., the average
number of connections per neuron  given by $cN$. In order to allow for 
variable symmetry as well, we define a joint-probability distribution
for $i < j$ ($c_{ii}=1$)
\begin{align}
\label{toch}
\textrm{P}[(c_{ij},c_{ji})=(x,y)]=
    &\quad\left(c_1^2+\frac{u+2v+w}{4}\right)
     \delta_{x,1}\delta_{y,1}+\left(c_2^2+\frac{u-2v+w}{4}\right)
      \delta_{x,-1}\delta_{y,-1}
                        \nonumber\\
    &+\left(c_1(1-c_1-c_2)-\frac{v+w}{2}\right)
   \left(\delta_{x,1}\delta_{y,0}+\delta_{x,0}\delta_{y,1}\right)
                          \nonumber\\
    &+\left(c_2(1-c_1-c_2)+\frac{v-w}{2}\right)
   \left(\delta_{x,0}\delta_{y,-1}+\delta_{x,-1}\delta_{y,0}\right)
                         \nonumber\\
    &+\left(c_1c_2-\frac{u-w}{4}\right)
   \left(\delta_{x,1}\delta_{y,-1}+\delta_{x,-1}\delta_{y,1}\right)
                        \nonumber\\
     &+\left((1-c_1-c_2)^2+w\right)\delta_{x,0}\delta_{y,0}
   \end{align}
with
\begin{align}
   u&=\langle c_{ij}c_{ji}\rangle-\langle c_{ij}\rangle\langle
         c_{ji}\rangle=\langle c_{ij}c_{ji}\rangle-(c_1-c_2)^2
                  \nonumber\\
   v&=\langle c_{ij}^2c_{ji}\rangle-\langle c_{ij}^2\rangle\langle
         c_{ji}\rangle=\langle c_{ij}^2c_{ji}\rangle-c(c_1-c_2)
                     \nonumber\\
   w&=\langle c_{ij}^2c_{ji}^2\rangle-\langle c_{ij}^2\rangle\langle
   c_{ji}^2\rangle=\langle c_{ij}^2c_{ji}^2\rangle-c^2.
   \label{uvw}
\end{align}
We note that these expressions are symmetric under the change 
$i\leftrightarrow j$. \\

These distributions generalize the following cases of random dilution 
frequently discussed in the literature (see, e.g., \cite{B04})
\begin{itemize}
\item Symmetric dilution where $c_{ij}=c_{ji}$ (SD): Due to the symmetry, 
$u=c-(c_1-c_2)^2$, $v=(c_1-c_2)(1-c_1-c_2)$ and $w=c(1-c)$, yielding for
eq.~(\ref{toch})
\begin{equation}
     \label{SD}
\textrm{P}[(c_{ij},c_{ji})=(x,y)]=
     c_1\delta_{x,1}\delta_{y,1}+c_2\delta_{x,-1}\delta_{y,-1}+
        \left(1-c\right)\delta_{x,0}\delta_{y,0}
\end{equation} 
whereby in most cases $c_2$ is taken to be zero, indicating that there
are no anti-Hebbian couplings mixed in explicitly.
\item Asymmetric dilution with $c_{ij}\neq c_{ji}$ and $c_2=0$ (AD): 
 In this case $u=v=w=\langle c_{ji}c_{ji}\rangle-c^2$ and the 
 joint-probability distribution eq.~(\ref{toch}) becomes
\begin{align}
           \label{AD}
\textrm{P}[(c_{ij},c_{ji})=(x,y)]=
    &(u+c^2)\delta_{x,1}\delta_{y,1}
       +(c(1-c)-u)(\delta_{x,1}\delta_{y,0}+\delta_{x,0}\delta_{y,1})
                   \nonumber\\
     &+(1+u-2c+c^2)\delta_{x,0}\delta_{y,0}\, .
\end{align}
\end{itemize}

The meaning  of the variable dilution as introduced in
eq.(\ref{eq:JD})-(\ref{uvw}) can best be understood from the theory of 
random graphs with $N$ nodes and $p$  the probability that any two of 
them are connected (see, e.g., \cite{Bollobas}). The number of 
connections a node has for $N\rightarrow\infty$ (i.e., in the 
thermodynamic limit) goes to a Poisson distribution
\begin{equation}
P(n)=e^{\langle n\rangle}\frac{\langle n\rangle^n}{n!}
\end{equation}
telling us that $\left\langle n \right\rangle=pN$ is the average number
 of connections per node. 
In order to indicate  what range of dilution we allow for, we look
at the diameter of the random graph, $d$, i.e., the maximum 
distance between any pair of nodes. This diameter is concentrated around 
\begin{equation}
     d=\frac{\log{(N)}}{\log{(\langle n\rangle)}}=
                          \frac{\log{(N)}}{\log{(pN)}}\, .
\end{equation}
The diameter is clearly 1 in the case that there is an edge with
probability $1$ and, hence, we have a fully connected graph. The
diameter  diverges when $p \sim 1/N$. Given that $p$ scales with $N$ like 
$p\sim N^z$, several regimes containing different types of subgraphs
can be distinguished as a function of $z$ \cite{Bollobas}. In
particular, it can be shown that the precise point above which the system
becomes completely connected (such
that it is always possible to find a path between any two nodes) is  
$\langle n\rangle\geq\log{(N)}$. 

Looking at eq.~(\ref{ND})  we find that this describes precisely a random 
graph with $c=c_1 +c_2$ being the probability to have an edge.
It follows that the average number of connections per neuron is $cN$.
Since our $Q$-Ising network is taken to be a mean-field system
characterized by an extensive number of long-range interactions we need
to have that $\langle n\rangle = cN$  tends to infinity for $N \to
\infty$ and all $c$. This implies that
\begin{equation}
    \mbox{P}(|c_{ij}|=1)=\frac{\log{(N)}}{N}+c
\end{equation}
such that $cN=\log{(N)}$ for $c=0$ and the complete connectivity of our 
model is still guaranteed. In this way, the diameter of our network is $d=1$
for $\forall c>0$ and $d=\infty$ when $c=0$, and the average number of
connections is given by $\langle n\rangle=cN+\log{(N)}$. The limit 
$c \to 0$, i.e., the so-called extremely diluted limit 
has now a simple interpretation: each neuron has an infinite number of
neighboring neurons but such that the average distance
between any two neurons tends to infinity. All this is graphically
illustrated in figure 1. 
\begin{figure}[h]
\begin{center}
\includegraphics[scale=0.6]{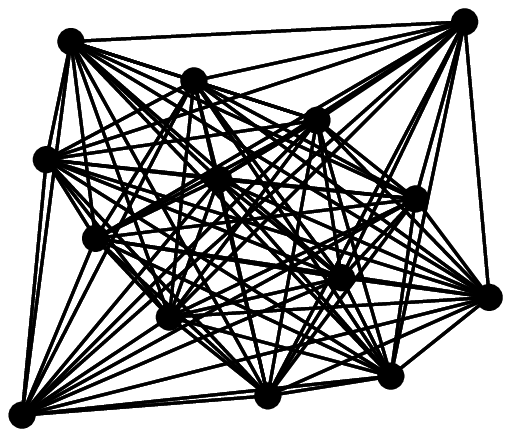}
$\quad$$\quad$\includegraphics[scale=0.6]{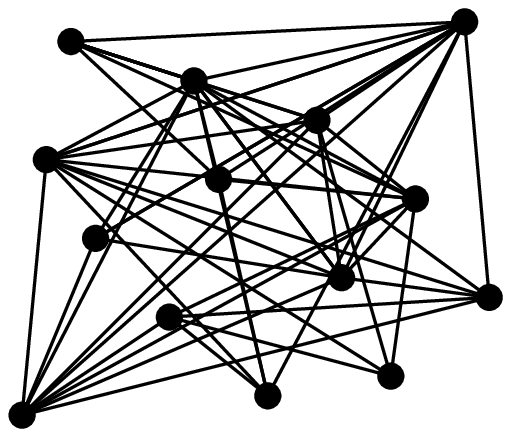}\\
\vspace{0.6cm}
\includegraphics[scale=0.6]{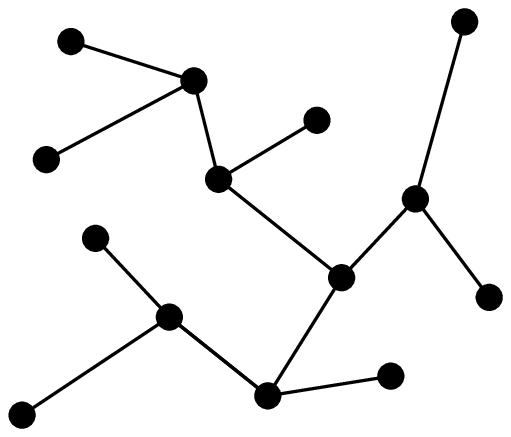}
$\quad$$\quad$\includegraphics[scale=0.6]{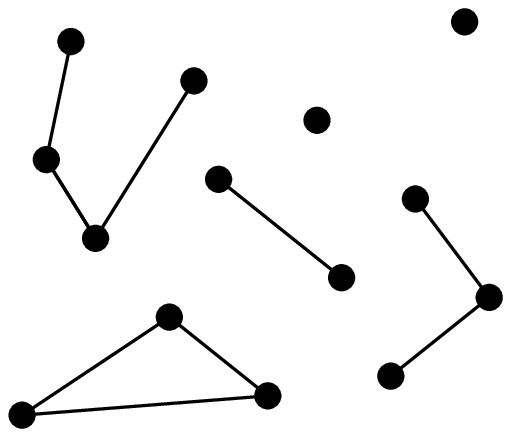}
\caption{Effect of dilution. Top left: $c=1$ (fully connected); top
  right: $0<c<1$ (some couplings are cut); bottom left: $c=0$ (extreme
dilution, tree-like structures); and bottom right:  not all sites are 
connected (= finite connectivity).}
\end{center}
\end{figure}

We remark that in this way we can understand that eq.~(\ref{eq:JD}) has
a well-defined meaning. There is a factor $cN$ in the denominator which
always tends to infinity in the thermodynamic limit $N \to \infty$, 
whatever the value of $c$. Finally, we note
that a distribution for the $c_{ij}$ can be chosen that does not support
complete connectivity (see figure 1 bottom right), e.g., 
by taking $c=\tilde{c}/N$, with $\tilde{c}$ a fixed
number independent of $N$. The system  then consists of disjoint
clusters of different sizes (=finite connectivity) \cite{WC}. These models
are much more difficult to handle and are outside the scope of the present
work.

\section{Symmetric dilution and statics of the $Q$-Ising model}
\label{sec3}
It is well-known that networks with symmetric couplings $J_{ij}$ obey
the detailed balance principle. Systems with detailed balance can be
described by standard equilibrium statistical mechanics making use of a
Hamiltonian. The $Q$-Ising neural network we have defined in Section 2
is of a mean-field type with couplings (\ref{eq:JD}) of infinite range and
restricting ourselves to symmetric dilution $c_{ij}=c_{ji}$
(cfr., eq.~(\ref{SD})) the couplings $J_{ij}^c$ remain symmetric. The
long time behavior of this network for sequential updating of the neurons
is then governed by the following Hamiltonian 
\begin{equation}   
  H=- \frac{1}{2} \sum_{i \neq j} J_{ij}^c \sigma_i \sigma_j 
     + b \sum_i \sigma_i^2 \, .
\end{equation}
For parallel updating of the neurons a Hamiltonian is defined in terms
of a two-spin representation \cite{Peretto} as  
\begin{equation}   
  H=- \sum_{ij} J_{ij}^c \sigma_i \tau_j 
     + b \sum_i (\sigma_i^2 + \tau_i^2) \, .
\end{equation}
The thermodynamic properties of the system are then determined from the free
energy using the standard techniques of replica mean-field theory
\cite{VB,MPV87}. It is outside the scope of this contribution to go
into the very details of these techniques but we indicate how the dilution
is going to affect the calculations and the results. 

The free energy is given as the logarithm of the partition function
averaged over the disorder. This average is done using the replica
technique such that we can write  
\begin{equation}
  Z^n({\bxi})=\underset{\bsigma}{\mbox{Tr}}
    \exp{\left(-\beta\sum_{\alpha=0}^nH({\bxi},{\bsigma}^\alpha)\right)}
\end{equation}
where $H({\bxi},{\bsigma}^\alpha)$ is the Hamiltonian of each replica of the
system given above and $\alpha$ is the replica index. Next, we first average
this partition function over the dilution. At this point we remark 
that we consider the case of sequential updating. Parallel updating does
not contain any additional difficulties \cite{BEVp}. Furthermore, we do
not write explicitly the term proportional to $b$ in the Hamiltonian
since it does not contain the $c_{ij}$ such that it can be inserted at
any point of the calculation.
Due to the central limit theorem we can easily see that  
$J_{ij}^c \sim \mathcal{O}((cN)^{-1/2})$, so that we can expand 
\begin{align}
   Z^n&=\underset{{\bsigma}}{\mbox{Tr}}
                  \exp{\left(\frac{\beta}{2}\sum_\alpha
     \sum_{i,j\neq i}J_{ij}^c\sigma_i^\alpha\sigma_j^\alpha\right)}
                  \nonumber\\
     &=\underset{{\bsigma}}{\mbox{Tr}}
        \prod_{i<j}\left[1+\beta\sum_\alpha J_{ij}^c
     \sigma_i^\alpha\sigma_j^\alpha+\frac{\beta^2}{2}(J_{ij}^c)^2
     \left(\sum_\alpha\sigma_i^\alpha\sigma_j^\alpha\right)^2+
         \mathcal{O}((cN)^{-3/2}) \right] \, .
\end{align}
The average over the dilution variables $c_{ij}$ with $i<j$ denoted by
$\left\langle \cdot \right\rangle_{\bf c}$ is then straightforward
\begin{align}
\langle Z^n\rangle_{{\bf c}}&=\underset{{\bsigma}}{\mbox{Tr}}
     \prod_{i<j}\left[1+\beta\,\,\frac{c_1-c_2}{c}\sum_\alpha J_{ij}
     \sigma_i^\alpha\sigma_j^\alpha+\frac{\beta^2}{2c}(J_{ij})^2
   \left(\sum_\alpha\sigma_i^\alpha\sigma_j^\alpha\right)^2+
       \mathcal{O}((cN)^{-3/2}) \right]
                      \nonumber\\
   &\simeq\underset{{\bsigma}}{\mbox{Tr}}
     \exp{\left[\frac{\beta}{2}\,\,\frac{c_1-c_2}{c}
        \sum_\alpha\sum_{i,j\neq i}J_{ij}
  \sigma_i^\alpha\sigma_j^\alpha
             +\frac{\beta^2}{4}\,\,\frac{c-(c_1-c_2)^2}{c^2}
        \sum_{i,j \neq i}(J_{ij})^2
   \left(\sum_\alpha\sigma_i^\alpha\sigma_j^\alpha\right)^2
         \right]}
\end{align}
where we have used that $J_{ij}\sim \mathcal{O}(N^{-1/2})$. Finally, noting
that $(J_{ij})^2\rightarrow (\alpha c)/{N}+\mathcal{O}((cN)^{-2})$
as $N\rightarrow\infty$, with $\alpha={p}/(cN)$ the finite loading 
capacity, we write the replicated partition function
averaged over the dilution as
\begin{equation}
\langle Z^n\rangle_{{\bf c}}=\underset{{\bsigma}}{\mbox{Tr}}
    \exp{\left[\frac{\beta}{2}\,\,\frac{c_1-c_2}{c}
       \sum_\alpha\sum_{i,j\neq i}J_{ij}\sigma_i^\alpha\sigma_j^\alpha
       +\frac{\beta^2\alpha}{4N}\,\,\frac{c-(c_1-c_2)^2}{c}\sum_{i,j\neq i}
      \left(\sum_\alpha\sigma_i^\alpha\sigma_j^\alpha\right)^2\right]}
                          \, .
\end{equation}
Using a Hubbard-Stratonovich transformation this can be further expressed
as
\begin{equation}
\langle Z^n\rangle_{{\bf c}}=\left\langle
    \underset{{\bsigma}}{\mbox{Tr}}
    \exp{\left[\frac{\beta}{2}\sum_{\alpha}\sum_{i,j\neq i}
    \left(\frac{c_1-c_2}{c}J_{ij}+d_{ij}\right)
            \sigma_i^\alpha\sigma_j^\alpha\right]}
    \right\rangle_{{\bf d}}
    \label{dilav}
\end{equation}
with $\left\langle \cdot \right\rangle_{\bf d}$ indicating the average over 
the $d_{ij}$, a set of i.i.d.r.v. for 
$i<j$ and symmetric with $d_{ii}=0$. They obey a gaussian distribution with
 mean $\langle d_{ij}\rangle=0$ and variance 
$\langle d_{ij}^2\rangle=\alpha s/(cN)$ with $s=c-(c_1-c_2)^2$. 
This shows that the symmetric dilution (\ref{SD}), being a form of
multiplicative noise, introduces an effective Hamiltonian where the 
learning rule now contains an additive noise term plus a scaling of the
Hebbian part
\begin{equation}
J_{ij}^c=J_{ij}\frac{c_{ij}}{c}\quad\rightarrow\quad
                    J_{ij}^c=\frac{c_1-c_2}{c}J_{ij}+d_{ij}\,.
\label{JSTE}
\end{equation}
The scaling factor expresses the influence of explicitly allowing an
average amount $c_2= \textrm{P}[c_{ij}=-1]$ of anti-Hebbian couplings. 
For $c_2=0$ the scaling term is $1$ and in this case the expression 
(\ref{JSTE}) agrees with the results of Sompolinsky \cite{Som86} for the
Hopfield model $Q=2$ and with the results of Theumann and 
Erichsen \cite{TE01} for the symmetrically diluted $Q$-Ising model. 

An analogous  calculation can be performed for parallel updating of the
neurons leading precisely to eq.~(\ref{dilav}) with $\sigma_j^\alpha$ 
replaced by $\tau_j^\alpha$ and with the factor $1/2$ removed.
In all these calculations we have only used the first and  second moments of
the probability distribution for the $c_{ij}$. This is due to the
mean-field character of the network we have treated. Therefore, one
can easily extend this result to any (symmetric) multiplicative noise 
such that
\begin{equation}
     \label{eq1}
   J_{ij}^m=J_{ij}\frac{\eta_{ij}}{\eta}\quad\rightarrow\quad
   J_{ij}^a=\frac{\sqrt{\eta-s}}{\eta}J_{ij}+
                      \sqrt{\frac{\alpha s}{\eta N}}d'_{ij},
\end{equation}
with obvious meaning of the superscripts $m$ and $a$, with
$d'_{ij}=\mathcal{N}(0,1)$ a gaussian with mean zero and
variance $1$, and with $\eta=\langle\eta_{ij}^2\rangle$  and
$s=\eta-\langle\eta_{ij}\rangle^2$ characterizing the multiplicative noise. 

To close this Section we remark that when we are interested in the
further calculation of the free energy, e.g., in order to obtain the
equilibrium fixed-point equations and the loading capacity we have to 
average the partition function (\ref{dilav}) over the pattern 
distribution employing the standard techniques. We refer to the
literature for the final results \cite{TE01,BEVp}.

%\newpage
\section{General dilution and dynamics of the $Q$-Ising model}
\label{sec4}
For asymmetric dilution (recall, e.g., (\ref{AD})) the system we have
defined does not obey detailed balance. In this case we have to resort
to techniques used in non-equilibrium statistical mechanics. The method
we use to study the effect of general dilution in the dynamics is the 
generating functional approach \cite{C01,Dom78}. 

The idea of this approach is to look at the probability to find a
certain microscopic path in time. The basic tool to study the statistics
of these paths is the generating function
\begin{equation}
\big\langle\big\langle Z[{\bf \Phi}]\big\rangle_{{\bf c}}
                       \big\rangle_{{\bxi}}
          =\sum_{{\bsigma}(0)}...\sum_{{\bsigma}(t)}
    \exp{\left(-i\sum_{s=0}^t\sum_{i=1}^N(\phi_i(s)\sigma_i(s))\right)}
      \big\langle\big\langle \mbox{P}[{\bsigma}(0)...{\bsigma}(t)]
           \big\rangle_{{\bf c}}\big\rangle_{{\bxi}}
\end{equation}
where  the dilution and pattern averages are denoted by 
$\big\langle\big\langle \cdot \big\rangle_{{\bf c}}
\big\rangle_{{\bxi}}$ and where $\mbox{P}[{\bsigma}(0)...{\bsigma}(t)]$ 
is the probability to have a certain path in phase space
\begin{equation}
\mbox{P}[{\bsigma}(0)...{\bsigma}(t)] =\mbox{P}[{\bsigma}(0)]
\prod_{t'=0}^{t-1}W[{\bsigma}(t')|{\bsigma}(t'-1)]
\end{equation}
with $ W[{\bsigma}(t')|{\bsigma}(t'-1)]$ the transition probabilities
from ${\bsigma}(t'-1)$ to ${\bsigma}(t')$. For the $Q$-Ising network with
parallel updating they read
\begin{equation}
W[{\bsigma}(t')|{\bsigma}(t'-1)]=
  \prod_{i=1}^N\frac{\exp{\left(\beta\sigma_i(t')\sum_j
        J_{ij}^c\sigma_j(t'-1)- \beta b\sigma_i^2(t')\right)}}
        {\underset{\sigma}{\textrm{Tr}}
     \exp{\left(\beta\sigma\sum_jJ_{ij}^c\sigma_j(t'-1)-\beta
                b\sigma^2 \right)}} \, .
\end{equation}
We remark again that sequential updating can be treated in an analogous 
way. Furthermore, we note that one can obtain all the order parameters 
of the system
through derivation of the generating function, e.g., the overlap between
the network configuration and an embedded pattern is given by
\begin{equation}
  m(t)=\frac{1}{A}\big\langle\big\langle \xi_i\sigma_i(t)
       \big\rangle_{{\bf c}}\big\rangle_{{\bxi}}=
   i \, \lim_{{\bf \Phi}\rightarrow 0}
   \frac{1}{A}\Big\langle\Big\langle \frac{\xi_i \,
     \partial  Z[{\bf \Phi}]}  {\partial \phi_i(t)}
       \Big\rangle_{{\bf c}}\Big\rangle_{{\bxi}} \, .
\end{equation}
Introducing the 
local fields  ${\bf h} =\{h_i(s)=\sum_jJ_{ij}^c\sigma_j(s)\}$ and their 
conjugates $\hat {\bf h}$,  we arrive at
\begin{align}
\big\langle\big\langle Z[{\bf \Phi}]\big\rangle_{{\bf c}}
       \big\rangle_{{\bxi}}&=
   \int\{d{\bf h}\}\{d\hat{{\bf h}}\}
   \sum_{{\bsigma}(0)}...\sum_{{\bsigma}(t)}
   \mbox{P}[{\bsigma}(0)]\prod_{s>0}^t\prod_i\mbox{P}[\sigma_i(s)|
       h_i(s-1),b]
                      \nonumber\\
  &=\exp{\left(N\mathcal{F}[{\bsigma},\hat{{\bf h}}]\right)}
      \prod_i\prod_{s=0}^t\exp{\left(i\hat{h}_i(s)h_i(s)-i\phi_i(s)
           \sigma_i(s)\right)}
\end{align}
where the disorder and dilution are put in one term 
\begin{equation}
\mathcal{F}[{\bsigma},\hat{{\bf h}}]=
   \frac{1}{N}\log{\left( \Bigg\langle\Bigg\langle
    \exp{\left[-i\sum_i\sum_{s=0}^t\hat{h}_i(s)
        \sum_jJ_{ij}^c\sigma_j(s)\right]}
               \Bigg\rangle_{{\bf c}}\Bigg\rangle_{{\bxi}}\right)} \, .
\end{equation}
We do not report the whole treatment of the generating function here but
refer to, e.g., \cite{C01} for more details on the method. We do
discuss in detail, however, the dilution average since this is precisely
the subject of our study.  

As in the statics  we use that the diluted couplings $J_{ij}^c$ are of order
$\mathcal{O}((cN)^{-1/2})$ and that the 
$J_{ij}\sim \mathcal{O}(N^{-1/2})$. We note that, in principle, diagonal
coupling terms usually taken to be of the form $J_{ii}=\alpha J_0$ can be 
present, but they are taken out separately
because they do not need to be averaged over. Introducing then 
$b_{ij}=\sum_s\hat{h}_i(s)\sigma_j(s)$ and using the fact that the
distribution for the  asymmetric dilution eq.~(\ref{toch}) is i.i.d.r.v. 
for $i<j$ we can write 
\begin{align}
    \bigg\langle\cdot\bigg\rangle_{{\bf c}}&=
     \prod_{i<j}\big\langle\exp{[-i(J_{ij}^cb_{ij}+J_{ji}^cb_{ji})]}
                 \Big\rangle_{(c_{ij},c_{ji})}
                   \nonumber\\
    &=\prod_{i<j}\Big\langle
     1-\frac{i}{c}J_{ij}\left(c_{ij}b_{ij}+c_{ji}b_{ji}\right)
        -\frac{1}{2c^2}J_{ij}^2\left(c_{ij}b_{ij}+c_{ji}b_{ji}\right)^2
        + \mathcal{O}((cN)^{-3/2})\Big\rangle_{(c_{ij},c_{ji})}
                 \nonumber\\
    &=\prod_{i,j\neq i}\exp{\left[-i\,\,\frac{c_1-c_2}{c}J_{ij}b_{ij}
      -\frac{s}{4c^2}J_{ij}^2(b_{ij}+b_{ji})^2-
        \frac{u-s}{2c^2}J_{ij}^2b_{ij}b_{ji}\right]}
                    \nonumber\\
     &=\prod_{i,j\neq i}\exp{\left[-i\,\,\frac{\sqrt{c-s}}{c}J_{ij}b_{ij}
          -\frac{s}{2c^2}J_{ij}^2(w_+b_{ij}+ w_-b_{ji})^2\right]}
\end{align}
where we have introduced a number of shorthand notations
\begin{equation}
s=c-(c_1-c_2)^2 \, , \qquad
w_{\pm}=\sqrt{\frac{1\pm\sqrt{1-\Gamma^2}}{2}} \, ,
        \qquad\Gamma\equiv\frac{u}{s}=
    \frac{\langle c_{ij}c_{ji}\rangle-\langle c_{ij}\rangle^2}
           {\langle c_{ij}^2\rangle-\langle c_{ij}\rangle^2}=
    \frac{\langle c_{ij}c_{ji}\rangle-(c_1-c_2)^2}{c-(c_1-c_2)^2}\, .
\end{equation}
We remark that the parameter $\Gamma$ is a measure for the symmetry in 
the $c_{ij}$, and  takes values in the range $[-1,1]$; $\Gamma=1$ is 
complete symmetry, $\Gamma=-1$ is complete antisymmetry. 
Finally, a Hubbard-Stratonovich transformation can be done as before 
leading to 
\begin{equation}
\bigg\langle\cdot\bigg\rangle_{{\bf c}}=\Bigg\langle
\prod_{i,j\neq i}\exp{\left[-i(\frac{c_1-c_2}{c}J_{ij}+[w_+d_{ij}+
w_-d_{ji}])b_{ij}\right]}
\Bigg\rangle_{{\bf d}}
\end{equation}
where the $d_{ij}$ are a set of i.i.d.r.v. for $i<j$, symmetric,
$d_{ij}=d_{ji}$,  and obeying a gaussian probability distribution with  
 $\langle d_{ij}\rangle=0$ and  $\langle d_{ij}^2\rangle=
sJ_{ij}^2/c^2$. This shows that also general asymmetric dilution can be 
written as additive noise in the learning rule with a scaling of the 
Hebbian part
\begin{equation}
J_{ij}^c=J_{ij}\frac{c_{ij}}{c}\quad\rightarrow\quad
J_{ij}^c=\frac{\sqrt{c-s}}{c}J_{ij}+
       \sqrt{\frac{\alpha s}{cN}}(w_+d'_{ij} + w_-d'_{ji}),
\end{equation}
where now $d'_{ij}=\mathcal{N}(0,1)$.
 We note explicitly that the parameters $v$ and $w$ 
(recall eq.~(\ref{uvw})) do not play a role in the
calculation. Only the mean $c$, the variance $s$ and
the covariance $u$ of the probability distribution for the random
dilution are important. Again, for  $c_2=0$ the scaling factor
expressing the explicit mixing in of anti-Hebbian couplings is $1$ and 
in the symmetric case (SD), $\Gamma=1$ in addition, such that the
additive noise reduces to $J_{ij}^c=J_{ij}+d_{ij}$ 
(see eq.~(\ref{JSTE})).

As in the statics, the calculation of the dynamics can be pursued by doing 
the average over the patterns and expressions for the overlap,
correlation functions and response functions can be obtained. This is
beyond the purpose of the present contribution and will be presented in
\cite{BEVp}. Finally, the remarks made before that sequential dynamics
can be treated similarly and that the effective dynamics is formally
the same, boil down to the fact that the explicit form
of the transition rates is not needed to derive the effective path
average. Only the initial conditions need to factorize over the site
index $i$ and this is a characteristic property of mean-field systems.   

Using the generating functional approach it is clear that the effect of 
random dilution on the learning rule in  neural networks based on other 
types of multi-state neurons \cite{B04}, e.g., Blume-Emery-Griffiths 
neurons, Potts  neurons, Ashkin-Teller neurons
can be examined in an analogous way.

%\newpage
\section{Concluding remarks}
\label{sec5}
In this work we have examined the effects of general random dilution,
which can be considered as a static disruption of the learning process
and, hence, as a form of multiplicative noise in the Hebbian learning
rule, on the statics and dynamics of $Q$-Ising multi-state neural
networks. A fraction of the couplings is explicitly taken to be 
anti-Hebbian.
Both sequential updating and parallel updating of the neurons
are allowed. 
\begin{figure}[h]
\begin{center}
\includegraphics[angle=270,scale=0.35]{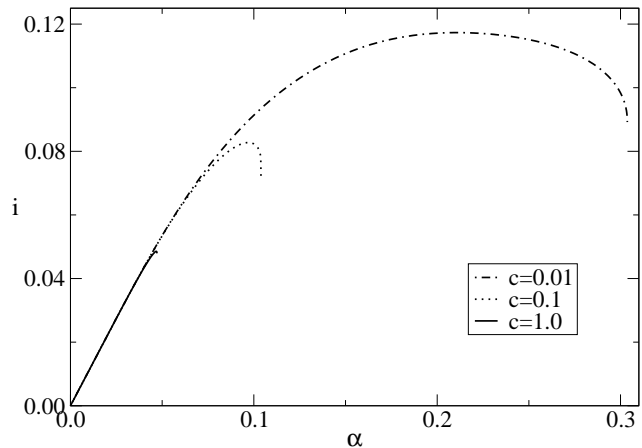}
\caption{ Information content $i$ as a function of the loading capacity
$\alpha$ for the $Q=3$ Ising model with uniform
patterns ($A=2/3$), $b=0.5$, $T=0$, and symmetric dilution with $c_2=0$
and $c_1=c=0.01$, $0.1$ and $1.0$.}
\end{center}
\end{figure}
It is shown, using replica mean-field theory, that for 
symmetric dilution the effect on the
learning rule appears as additive gaussian noise together with a 
scaling of the Hebbian part. This scaling is a measure for the average
amount of anti-Hebbian couplings and becomes $1$ when no such couplings
are present. This extends previous results in the 
literature. Moreover, for
general dilution, including asymmetric forms, a similar result is
obtained using the generating functional approach employed in studies of
the dynamics of disordered systems. The additive noise is determined as
a function of the mean, the variance and the covariance of the probability
distribution characterizing the dilution. We conjecture that this result
is valid for any network of mean-field type.

Although this is beyond the scope of the present work it is relevant to 
remark that it can be shown that the type of multi-state networks studied
here are robust against the interference of static noise coming from
random dilution (cfr., e.g., \cite{BEVp,Som86}) in the sense that the 
quality of the retrieval properties is affected very little, unless the 
amount of dilution is very high. As an illustration of this fact we show in
fig.~2 the information content, being the product of the loading
capacity and the mutual information, of a $Q=3$-Ising neural network with
parameters as indicated in the figure caption 
for several amounts of symmetric dilution $c$. For more details on this
we refer to \cite{BEVp}.

The fact that the effect of random dilution can be
expressed as additive noise in the learning rule makes the analytical
calculations on these networks easier and more transparent and can be of
help in the non-trivial numerical simulations of diluted systems.

%\newpage
\section{Acknowledgements}
We thank Rubem Erichsen Jr. for informative discussions. This work has
been supported in part by the Fund of Scientific Research,
Flanders-Belgium.

\end{document}